\documentclass[11pt]{article}
\usepackage{times}
\usepackage{sidecap}
\usepackage{geometry}
\usepackage{floatrow}
\usepackage{float}

\usepackage[
    colorlinks=true,
    linkcolor=black,
    citecolor=darkcyan,
    urlcolor=darkcyan
]{hyperref}
\usepackage{graphics,graphicx}
\usepackage{subcaption}
\usepackage{epsf,wrapfig}

\usepackage{natbib,bm}

\usepackage{paralist}
\let\olditem\item
\renewenvironment{thebibliography}[1]{%
  \section*{\refname}
  \let\par\relax
  \renewcommand{\item}[1][]{\olditem[\textbullet]}%
  \inparaenum}{\endinparaenum}

\usepackage{authblk}

\makeatletter
\newcommand{\printauthorlist}{%
  \begingroup
  \let\and\relax
  \@author
  \endgroup
}
\makeatother

\usepackage{etoolbox}

\newcounter{affiliation}
\newcommand{\affiliationlist}{}

\newcommand{\affiliation}[2]{%
  \refstepcounter{affiliation}%
  \label{aff:#1}%
  \gappto{\affiliationlist}{%
    \noindent\textsuperscript{\ref{aff:#1}}%
    \textit{#2}\par
  }%
}

\newcommand{\affmark}[1]{\textsuperscript{\ref{aff:#1}}}

\newcommand{\printaffiliations}{%
  \par\medskip
  {\footnotesize
  \section*{Author Affiliations}
  \affiliationlist
  }
}

\DeclareGraphicsExtensions{.jpg,.pdf,.png,.eps}

\usepackage{caption}
\usepackage{xcolor}

\usepackage{tcolorbox}

\bibpunct{(}{)}{;}{a}{}{,}

\usepackage{soul}

\definecolor{darkcyan}{rgb}{0.0, 0.0, 0.7}

\definecolor{boxcolor}{RGB}{193, 229, 255}
\definecolor{figboxcolor}{RGB}{193, 229, 255}
\definecolor{boxcolorgray}{RGB}{227, 227, 227}
\definecolor{darkcyan}{rgb}{0.0, 0.0, 0.7}

\sethlcolor{boxcolorgray}

\def\gtrsim{\lower 2pt \hbox{$\, \buildrel {\scriptstyle >}\over
{\scriptstyle \sim}\,$}}
\def\lesssim{\lower 2pt \hbox{$\, \buildrel {\scriptstyle <}\over
{\scriptstyle \sim}\,$}}

\def\approxlt{\lower.2em\hbox{$\buildrel < \over \sim$}}
\def\approxgt{\lower.2em\hbox{$\buildrel > \over \sim$}}

\def\arcsec{$^{\prime\prime}$}

\def\be{\begin{equation}}
\def\ee{\end{equation}}
\def\bea{\begin{eqnarray*}}
\def\eea{\end{eqnarray*}}

\def\arcsec{^{\prime\prime}}

\def\jref@jnl#1{{\rm#1}}
\def\aj{\jref@jnl{AJ}}			
\def\araa{\jref@jnl{ARA\&A}}		
\def\apj{\jref@jnl{ApJ}}			
\def\apjl{\jref@jnl{ApJ}}		
\def\apjs{\jref@jnl{ApJS}}		
\def\ao{\jref@jnl{Appl.~Opt.}}		
\def\apss{\jref@jnl{Ap\&SS}}		
\def\aap{\jref@jnl{A\&A}}		
\def\aapr{\jref@jnl{A\&A~Rev.}}		
\def\aaps{\jref@jnl{A\&AS}}		
\def\azh{\jref@jnl{AZh}}			
\def\baas{\jref@jnl{BAAS}}		
\def\jrasc{\jref@jnl{JRASC}}		
\def\memras{\jref@jnl{MmRAS}}		
\def\mnras{\jref@jnl{MNRAS}}		
\def\pra{\jref@jnl{Phys.~Rev.~A}}	
\def\prb{\jref@jnl{Phys.~Rev.~B}}	
\def\prc{\jref@jnl{Phys.~Rev.~C}}	
\def\prd{\jref@jnl{Phys.~Rev.~D}}	
\def\pre{\jref@jnl{Phys.~Rev.~E}}	
\def\prl{\jref@jnl{Phys.~Rev.~Lett.}}	
\def\pasp{\jref@jnl{PASP}}		
\def\pasj{\jref@jnl{PASJ}}		
\def\qjras{\jref@jnl{QJRAS}}		
\def\skytel{\jref@jnl{S\&T}}		
\def\solphys{\jref@jnl{Sol.~Phys.}}	
\def\sovast{\jref@jnl{Soviet~Ast.}}	
\def\ssr{\jref@jnl{Space~Sci.~Rev.}}	
\def\zap{\jref@jnl{ZAp}}			
\def\nat{\jref@jnl{Nature}}		
\def\iaucirc{\jref@jnl{IAU~Circ.}}
\def\aplett{\jref@jnl{Astrophys.~Lett.}}
\def\apspr{\jref@jnl{Astrophys.~Space~Phys.~Res.}}
\def\bain{\jref@jnl{Bull.~Astron.~Inst.~Netherlands}}
\def\fcp{\jref@jnl{Fund.~Cosmic~Phys.}}
\def\gca{\jref@jnl{Geochim.~Cosmochim.~Acta}}
\def\grl{\jref@jnl{Geophys.~Res.~Lett.}}
\def\jcp{\jref@jnl{J.~Chem.~Phys.}}	
\def\jgr{\jref@jnl{J.~Geophys.~Res.}}	
\def\jqsrt{\jref@jnl{J.~Quant.~Spec.~Radiat.~Transf.}}
\def\memsai{\jref@jnl{Mem.~Soc.~Astron.~Italiana}}
\def\nphysa{\jref@jnl{Nucl.~Phys.~A}}
\def\physrep{\jref@jnl{Phys.~Rep.}}
\def\physscr{\jref@jnl{Phys.~Scr}}
\def\planss{\jref@jnl{Planet.~Space~Sci.}}
\def\procspie{\jref@jnl{Proc.~SPIE}}

\def\sun{\hbox{$\odot$}}

\def\arcsec{\hbox{$^{\prime\prime}$}}

\def\ion#1#2{#1$\;${\small\rm\@Roman{#2}}\relax}

\newcount\lecurrentfam
\def\LaTeX{\lecurrentfam=\the\fam \leavevmode L\raise.42ex
\hbox{$\fam\lecurrentfam\scriptstyle\kern-.3em A$}\kern-.15em\TeX}
\def\sizrpt{
(\fontname\the\font): em=\the\fontdimen6\font, ex=\the\fontdimen5\font
\typeout{
(\fontname\the\font): em=\the\fontdimen6\font, ex=\the\fontdimen5\font
}}
\definecolor{dark}{rgb}{ 0.34, 0.09, 0.49}

\usepackage{fancyhdr,url}
\usepackage{transparent}
\usepackage{tablefootnote}
\usepackage{mathpazo}

\affiliation{udp}
{Instituto de Estudios Astrof\'isicos, Universidad Diego Portales,
Av. Ej\'ercito 441, Santiago, Chile}

\affiliation{mingal}{Millenium Nucleus for Galaxies (MINGAL)}

\affiliation{leiden}
{Leiden Observatory, Leiden University, NL-2300 RA Leiden, The Netherlands}

\affiliation{puc}
{Instituto de Astrof\'isica, Pontificia Universidad Cat\'olica de Chile,
Av. Vicu\~na Mackenna 4860, Santiago, Chile}

\affiliation{mpia}
{Max-Planck-Institut f\"ur Astronomie,
K\"onigstuhl 17, 69117 Heidelberg, Germany}

\affiliation{iram}
{Institut de Radioastronomie Millimétrique (IRAM), 300 rue de la Piscine, 38400 Saint-Martin-d’Hères, France}

\affiliation{iap}
{Sorbonne Universit\'e, UPMC Paris 6 and CNRS,
Institut d'Astrophysique de Paris, 98 bis boulevard Arago,
75014 Paris, France}

\affiliation{inafbologna}
{INAF--Osservatorio di Astrofisica e Scienza dello Spazio,
Via P. Gobetti 93/3, 40129 Bologna, Italy}

\affiliation{forth}
{Institute of Astrophysics, Foundation for Research and Technology–Hellas (FORTH), Heraklion 70013, Greece}

\affiliation{caltech}
{Caltech/IPAC, 1200 E. California Blvd., Pasadena, CA 91125, USA}

\affiliation{strasbourg}
{Université de Strasbourg, CNRS, Observatoire astronomique de Strasbourg, UMR 7550, F-67000 Strasbourg, France}

\affiliation{cab}
{Centro de Astrobiolog\'ia (CAB), CSIC-INTA, Spain}

\affiliation{cambridge}
{Kavli Institute for Cosmology, University of Cambridge, Madingley Road, Cambridge CB3 0HA, UK}

\affiliation{anu}
{Research School of Astronomy and Astrophysics,
Australian National University, Canberra, ACT 2611, Australia}

\affiliation{atacama}
{Instituto de Astronom\'ia y Ciencias Planetarias,
Universidad de Atacama, Copayapu 485, 1531772 Copiap\'o, Chile}

\affiliation{sissa}
{SISSA, Via Bonomea 265, 34136 Trieste, Italy}

\affiliation{purplemountain}{Purple Mountain Observatory, Chinese Academy of Sciences, 10 Yuanhua Road, Nanjing 210023, China}

\affiliation{chinaradiolab}{State Key Laboratory of Radio Astronomy and Technology, Purple Mountain Observatory, Chinese Academy of Sciences, 10 Yuanhua Road, Nanjing 210023, China}

\affiliation{inaftrieste}
{INAF--Osservatorio Astronomico di Trieste,
Via G. B. Tiepolo 11, 34143 Trieste, Italy}

\affiliation{michigan}
{Department of Astronomy, University of Michigan,
1085 South University Avenue, Ann Arbor, MI 48109, USA}

\affiliation{ucly}
{P\^ole Scientifique, Universit\'e Catholique de Lyon--Universit\'e de Lyon,
10 place des Archives, 69002 Lyon, France}

\affiliation{cologne}
{I. Physikalisches Institut, University of Cologne,
Z\"ulpicher Strasse 77, 50937 Cologne, Germany}

\affiliation{mpifr}
{Max-Planck-Institut f\"ur Radioastronomie,
Auf dem H\"ugel 69, 53121 Bonn, Germany}

\affiliation{eso}
{European Southern Observatory, Karl-Schwarzschild-Str. 2,
85748 Garching bei M\"unchen, Germany}

\affiliation{lancaster}
{Physics Department, Lancaster University, Lancaster LA1 4YB, UK}

\newcommand{\affmarks}[2]{%
  \textsuperscript{\ref{aff:#1},\ref{aff:#2}}%
}

\begin{document}

\noindent
\begin{raggedright}
{\Huge\bfseries Unveiling the Baryon Cycle Through Cosmic Time\\[0.5em]
\Large 
}\end{raggedright}
\noindent


\noindent
{\large
Manuel Aravena\affmarks{udp}{mingal},
Leindert A. Boogaard\affmark{leiden},
Jorge Gonz\'alez-L\'opez\affmarks{puc}{mingal},
Ryan Keenan\affmark{mpia},
Chiara Circosta\affmark{iram},
Pierre Cox\affmark{iap},
Roberto Decarli\affmark{inafbologna},
Tanio D\'iaz-Santos\affmark{forth},
Andreas Faisst\affmark{caltech},
Rom\'an Fern\'andez-Aranda\affmark{cab},
Jonathan Freundlich\affmark{strasbourg},
Carlotta Gruppioni\affmark{inafbologna},
Mahmoud Hamed\affmark{cab},
Gareth Jones\affmark{cambridge},
Melanie Kaasinen\affmark{anu},
Adarsh Kumar\affmark{atacama},
Ivanna Langan\affmark{cab},
Andrea Lapi\affmark{sissa},
Daizhong Liu\affmarks{purplemountain}{chinaradiolab},
Umberto Maio\affmark{inaftrieste},
Chiara Mazzucchelli\affmark{udp},
Dieu Nguyen\affmark{michigan},
Mamta Pandey-Pommier\affmark{ucly},
Prachi Prajapati\affmarks{cologne}{mpifr},
Gerg\"{o} Popping\affmark{eso},
Dominik Riechers\affmark{cologne},
Matus Rybak\affmark{leiden},
Irene Shivaei\affmark{cab},
Lucie Rowland\affmark{leiden},
Livia Vallini\affmark{inafbologna},
Julie Wardlow\affmark{lancaster}
}


\vspace{1em}
\emph{Affiliations can be found at the end of the manuscript, after the references.}
\vspace{1em}

\begin{tcolorbox}[boxsep=3pt,left=0pt,right=0pt,top=1pt,bottom=1pt,colframe=boxcolor,colback=boxcolor]
\justify
 
\textbf{Abstract:}
Galaxy evolution is governed by the baryon cycle: gas accreted from the cosmic web and circumgalactic medium condenses into molecular gas, forms stars, and is redistributed by stellar and AGN feedback. Tracing the molecular phase is therefore essential to understanding how galaxies acquire, consume, and eject their baryons. Yet current molecular-gas surveys remain limited by sensitivity, area, cosmic variance, and spectral bandwidth.

ALMA2040 would combine a $\sim10\times$ sensitivity improvement with $\gtrsim512$~GHz simultaneous frequency coverage, delivering $\sim100\times$ faster mapping at fixed depth and a further $\gtrsim25\times$ gain in instantaneous bandwidth. Together, these advances would increase spectroscopic survey speed by orders of magnitude, enabling tiered ``wedding-cake'' strategies: an ultra-deep $\sim4$--5~arcmin$^2$ field probing faint and high-redshift reservoirs, a $\sim$0.25 deg$^2$ survey reaching molecular-line sensitivities comparable to current ultra-deep fields over a much larger cosmological volume, and thousands of independent pointings minimizing cosmic variance.

Broad spectral coverage would simultaneously detect multiple CO transitions, [C\,I], H$_2$O, feedback tracers, far-IR fine-structure lines, and dust continuum, constraining molecular gas, ISM conditions, obscured star formation, and gas flows within the same observations. Every targeted observation would also become a molecular deep field, extending these measurements to serendipitous galaxies. Together with future H\,I, optical, and near-IR surveys, ALMA2040 would enable a population-resolved view of the baryon cycle across cosmic time and stringent tests of galaxy formation models.
\end{tcolorbox}

\newpage{}

\section{Scientific context}

\subsection{The baryon cycle as a fundamental problem in galaxy evolution}

Galaxies grow through a continuous baryon cycle in which gas is accreted from the intergalactic and circumgalactic media (IGM and CGM), settles into atomic reservoirs, becomes molecular, and ultimately forms stars. Stellar evolution and accreting supermassive black holes subsequently return energy, momentum, metals, and gas to the interstellar medium (ISM) and CGM through feedback and outflows \citep{Peroux2020, Walter2020}. Understanding galaxy evolution therefore requires an inventory not only of stars and star formation, but of the different baryon phases and the rates at which material flows between them.

\begin{figure}[ht]
    \centering
    \includegraphics[width=0.7\linewidth]{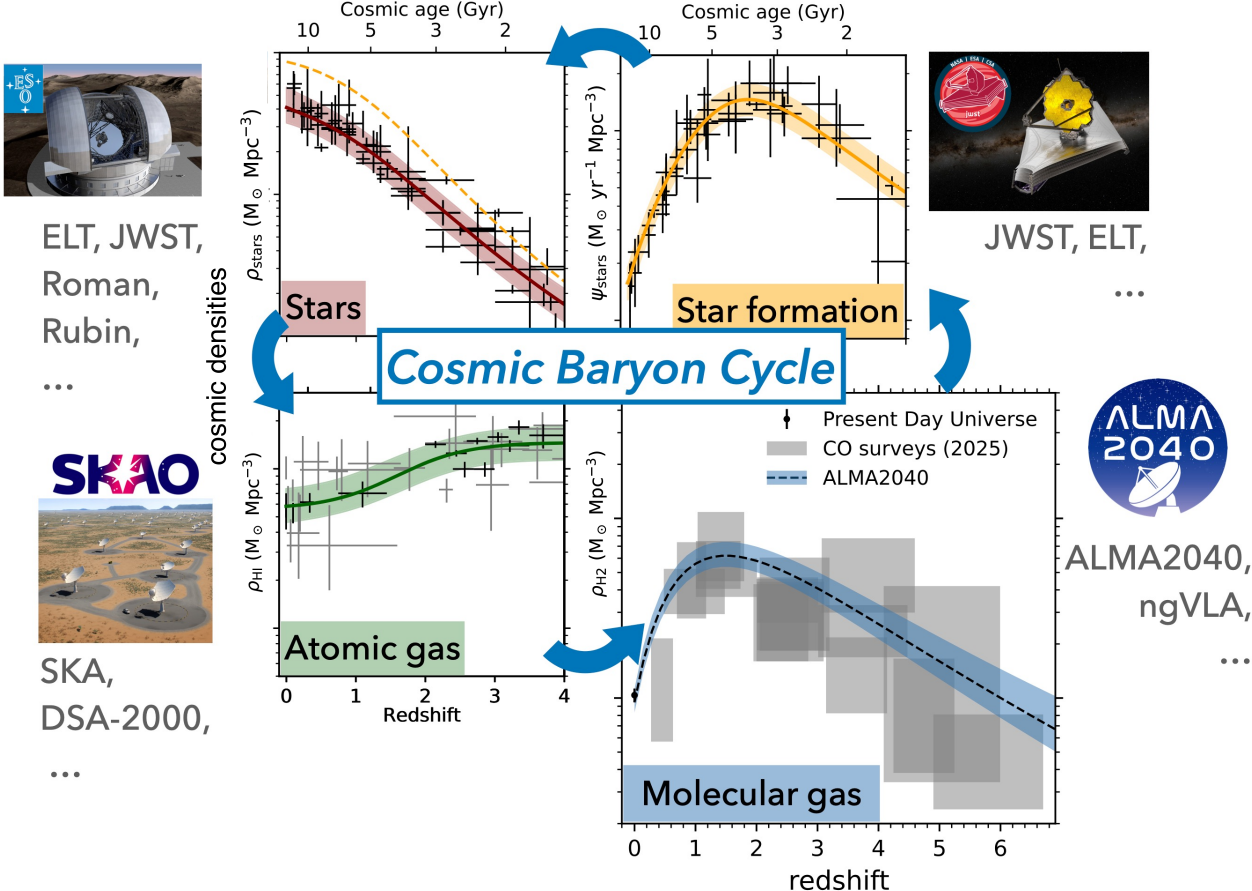}
    \caption{Redshift evolution of the cosmic SFR, stellar mass, atomic gas ($\rho_{\rm HI}$), and molecular gas ($\rho_{\rm H_2}$) densities, adapted from \citet{Walter2020}. Solid lines show the best-fit evolutionary trends, with shaded regions indicating the $1\sigma$ uncertainties. 
    The blue region illustrates the projected constraints achievable with ALMA2040 (see Fig.~\ref{fig:cosmicdensity} for more detailed predictions).}
    \label{fig:cosmicevolution}
\end{figure}

The first empirical cosmic baryon inventories have become possible only recently (Fig.~\ref{fig:cosmicevolution}). Measurements of the cosmic H\,I, H$_2$, stellar-mass, and star formation rate (SFR) densities show that the molecular component evolves strongly and is closely linked to the history of star formation \citep[e.g.,][]{Walter2020, Bollo2025}. The cosmic molecular gas density, $\rho_{\rm H_2}$, rises toward $z\sim1$--2 and decreases by about one dex toward both lower and higher redshift \citep[e.g.,][]{Decarli2019, Decarli2020, Riechers2019, Boogaard2023, Aravena2024, Bollo2025}.

In contrast, the cosmic H\,I density evolves comparatively weakly, while the stellar mass density continues to increase (Fig.~\ref{fig:cosmicevolution}). The molecular reservoir is therefore not a closed system: its depletion time, $t_{\rm dep}=M_{\rm mol}/{\rm SFR}$, is shorter than the Hubble time over the epochs where it has been measured, requiring continuous replenishment to explain the cumulative stellar mass observed at lower redshifts \citep{Walter2020}. The evolving ratios of molecular gas, atomic gas, and stellar mass are broadly consistent with ``bathtub'' or gas-regulator models in which accretion from the multiphase CGM continuously replenishes the material consumed by star formation and expelled by feedback \citep{Peroux2020}.

The challenge for the coming decades is therefore no longer simply to establish that high-redshift galaxies contain large molecular reservoirs. We must determine, across galaxy populations and cosmological volumes, \emph{which galaxies contain the baryons, in which phases they reside, how efficiently they form stars, and how rapidly feedback redistributes them}. This requires a qualitatively different experiment from most current studies of the cold ISM.

\subsection{The molecular Universe from targeted surveys: strengths and limitations}

Targeted molecular-gas surveys observe galaxies selected in advance through their redshifts and properties measured at other wavelengths. CO emission remains the most extensively calibrated tracer of molecular gas in galaxies \citep[e.g.][]{Carilli2013}. Over the last two decades, targeted programs such as the Plateau de Bure High-$z$ Blue Sequence Survey (PHIBSS and PHIBSS2), together with ALMA, NOEMA, and VLA studies, established scaling relations between molecular gas content, stellar mass, SFR, redshift, and offset from the star-forming main sequence \citep[e.g.][]{Daddi2010, Scoville2017, Liu2019, Tacconi2020}. These studies show that molecular gas fractions rise strongly toward $z\sim2$--3, while depletion times evolve more moderately.

Targeted samples, however, inherit the selection of their parent populations. Blind gas-selected surveys provide a complementary view. The ALMA Spectroscopic Survey in the Hubble Ultra Deep Field (ASPECS; see Section \ref{section:cosmicgascensus}), for example, identified CO-rich systems below the star-forming main sequence with substantial molecular reservoirs and long depletion times \citep{Aravena2019, Boogaard2019}, while its faint dust-selected population spans a broader range in star-formation activity than classical submillimeter-selected samples \citep{Aravena2020}. At $z\gtrsim4$, targeted [C\,II] surveys extend cold-gas studies to UV-selected galaxies, but inherit the selection of those populations and introduce additional uncertainty when [C\,II] is converted to molecular gas mass \citep{Zanella2018}. 

\subsection{Molecular deep fields and the cosmic gas census}
\label{section:cosmicgascensus}

Blind spectral-scan surveys have transformed molecular gas studies from targeted measurements into cosmological censuses. These experiments enabled direct measurements of CO luminosity functions and $\rho_{\rm H_2}$, establishing its rise toward cosmic noon and subsequent decline \citep[e.g.,][]{ Walter2016, Decarli2019, Decarli2020, Riechers2019, Boogaard2023}.

At the depths reached by current molecular deep fields, cosmic variance is a major limitation on population-level measurements because the surveyed volumes sample few independent structures and small samples retain substantial Poisson uncertainty. ASPECS covered only $\sim4.6$~arcmin$^2$, while the NOEMA HDF-North Survey covered twice that area but only at one-third of the depth \citep{Boogaard2023}.  Simulations show that such fields can bias the recovered CO luminosity function and $\rho_{\rm H_2}$ by tens of percent or more \citep{Popping2019, Keenan2020, Gkogkou2023}, indicating that surveys of order $\gtrsim50$--70~arcmin$^2$ are required before cosmic variance and Poisson noise become comparable for representative CO populations. Greater depth, however, remains essential for constraining the faint end of the population and extending direct measurements toward the earliest epochs \citep{Popping2026}.

A complementary solution is to distribute observations over many independent sightlines. ALMACAL constrained the CO luminosity function to $z\sim6$ using hundreds of calibrator fields and estimates field-to-field variance below $\sim5\%$ for the transitions considered \citep{Bollo2025}. These results motivate an ALMA2040 three-tier strategy: ultra-deep fields to probe the faint and highest-redshift populations, wide contiguous surveys to sample environments and large-scale structure, and distributed pointings to maximize effective area and minimize cosmic variance.

\begin{figure}
    \centering
    \includegraphics[width=0.7\linewidth]{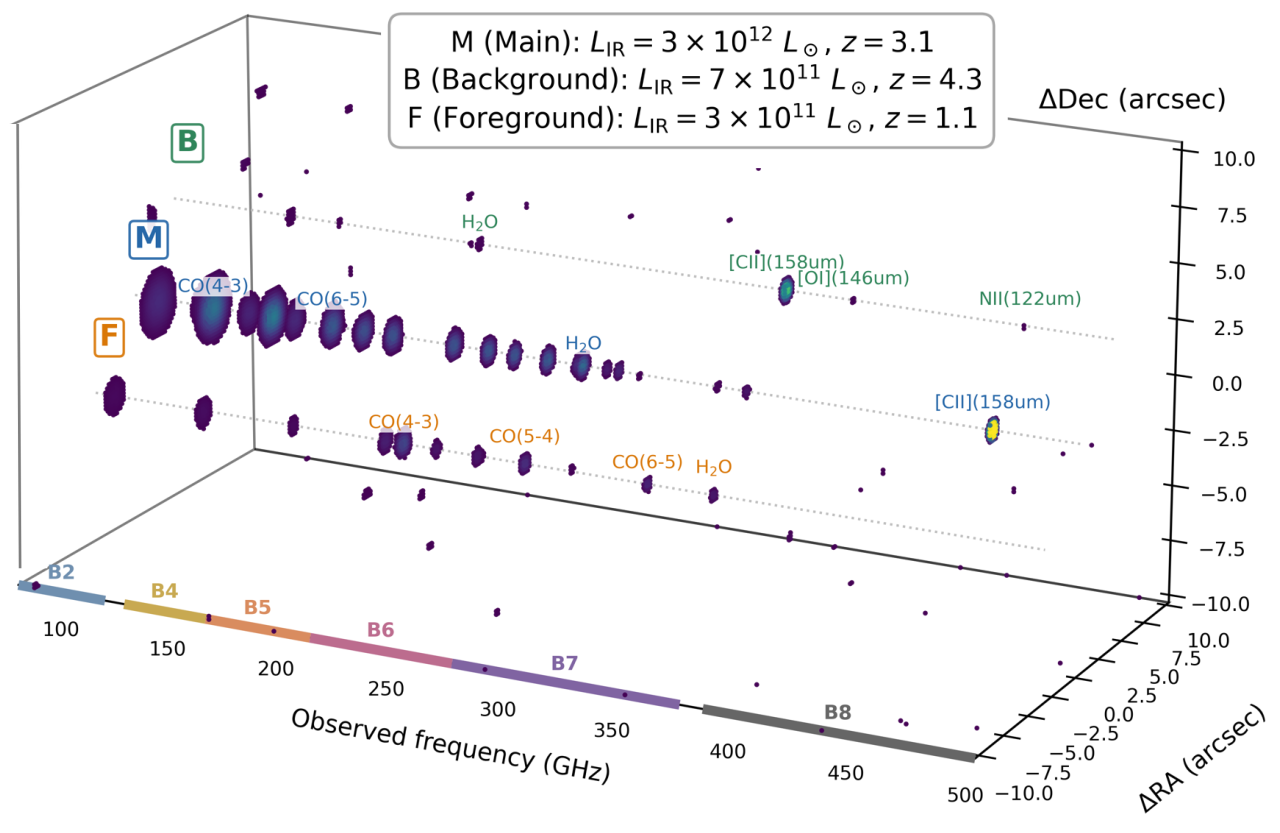}
    \caption{Simulated 10-min ALMA2040 spectral tomography of three star-forming galaxies within a single field of view: the main target (M), a background (B), and a foreground source (F). Line-emitting voxels with $S/N>5$ are shown as a function of observed frequency and sky position, with lower-significance emission masked. Prominent transitions and corresponding ALMA bands are indicated. For clarity, only Bands~2--8 are shown.}
    \label{fig:3dtomography}
\end{figure}

\section{Science case for ALMA2040}

\subsection{From deeper observations to a molecular gas survey machine}

A tenfold improvement in point-source sensitivity reduces the integration
required at fixed depth by a factor of 100. This gain can be invested either in vastly greater survey area or in substantially greater depth. Wider surveys increase sample size, reduce cosmic variance, probe environments and large-scale structure; ultra-deep observations instead access intrinsically fainter molecular reservoirs and push direct gas measurements toward the earliest epochs.

The gain becomes even more powerful when combined with the ALMA2040 goal of $\gtrsim512$~GHz simultaneous frequency coverage. While ASPECS required five frequency settings to cover $84–115$ GHz in Band 3 and eight settings to cover $212–272$ GHz in Band 6, or a total bandwidth of $\sim90$ GHz combined, ALMA2040 could simultaneously cover several atmospheric windows, from 35 to 500 GHz through bands 1-8, providing an enormous redshift path and access to multiple molecular, atomic, and far-IR transitions in a single observation. Sensitivity and spectral grasp together would therefore increase spectroscopic survey speed by orders of magnitude and transform both dedicated surveys and targeted observations into wide-band molecular deep fields.

The ngVLA will be highly complementary to this capability. Its exceptional sensitivity to low-$J$ CO will provide powerful measurements of the cold, low-excitation molecular reservoir and gas dynamics, particularly at high redshift. ALMA2040 will simultaneously access a fundamentally different suite of diagnostics, including multiple CO transitions, [C\,I], dense-gas tracers, H$_2$O, OH/OH$^+$, the FIR dust SED, and redshifted fine-structure lines such as [C\,II], [O\,I], [O\,III], and [N\,II]. Together, ngVLA and ALMA2040 would connect the cold molecular reservoir to its physical state, star formation, and feedback, providing a substantially more complete view of the baryon cycle than either facility alone (see the accompanying white papers).


These capabilities naturally enable a ``wedding-cake'' strategy. For example, an ultra-deep $\sim4$--5~arcmin$^2$ field would constrain the faint end of the molecular-gas population and extend the census toward $z\gtrsim7$. A $\sim0.25$~deg$^2$ survey at current deep-field sensitivity would sample environments over nearly 190 times the ASPECS area, while thousands of targeted observations would provide a complementary census across independent sightlines, maximizing effective area and minimizing cosmic variance. Together, these tiers span the depth, luminosity, environment, and volume required for a population-resolved view of the baryon cycle. Their depths and predicted yields are quantified in Sect.~\ref{section:moleculardeepfields}.

\subsection{Every targeted observation becomes a molecular deep field}

The distributed tier arises naturally because the distinction between ``targeted observations'' and ``deep fields'' will start to disappear with ALMA2040. Surveys such as A3COSMOS and ALMACAL already demonstrate the scientific potential of observations obtained for other purposes, including a blind CO census across hundreds of independent sightlines \citep[e.g.,][]{Liu2019, Hamanowicz2023,Bollo2025}. \textit{ALMA2040 would make this capability intrinsic to essentially every sufficiently deep observation.}

\begin{figure}
    \centering
    \includegraphics[width=0.99\linewidth]{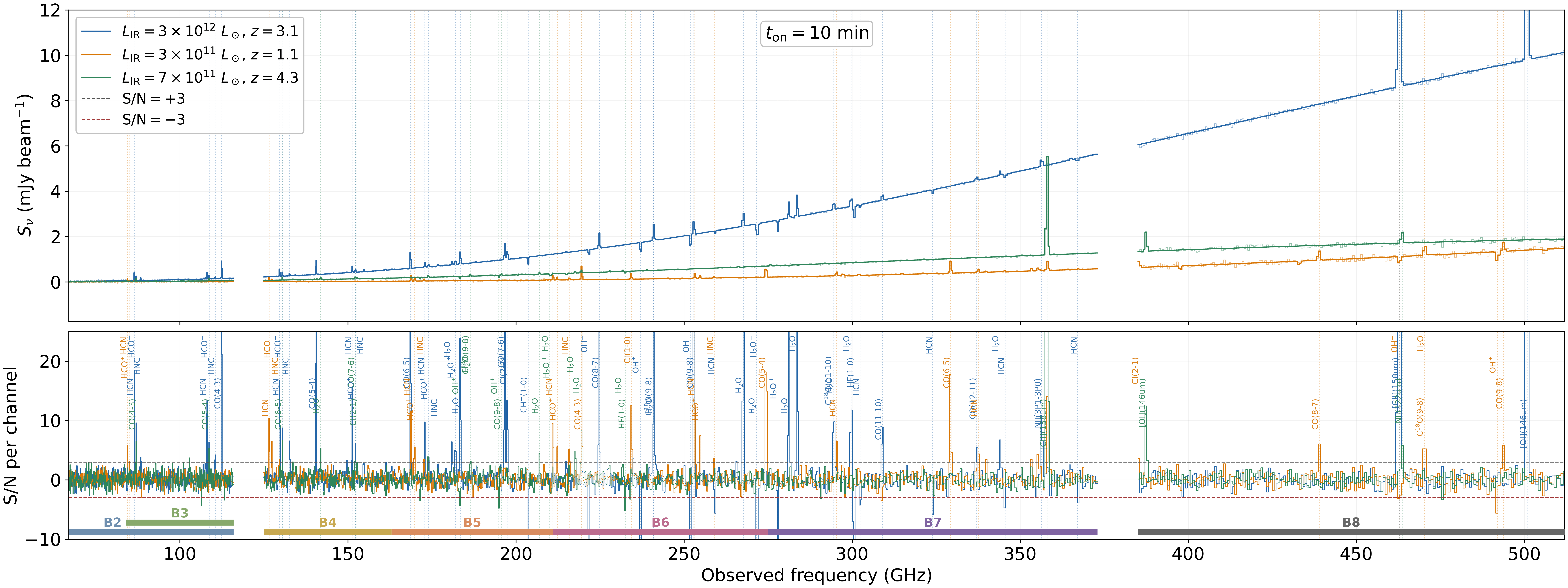}
    \caption{Simulated ALMA2040 spectra for the three dusty galaxies shown in Fig.~\ref{fig:3dtomography}, using $400~\mathrm{km\,s^{-1}}$ channels. The upper panel shows the mock observed and fitted line-plus-continuum spectra from a 10-min full-bandwidth ALMA2040 snapshot (see Sect.~\ref{sec:simulations}), while the lower panel presents the continuum-subtracted line $S/N$ per channel. Dashed lines mark $S/N=\pm3$, colored bars indicate the ALMA receiver bands, and vertical markers identify emission and absorption transitions expected to exceed $|S/N|=3$.}
    \label{fig:tomographyspectra}
\end{figure}

A pointing centered on a known galaxy would automatically survey the surrounding primary beam to depths comparable to present-day dedicated deep fields. The enormous simultaneous frequency coverage would probe a large redshift path for serendipitous emitters, while multiple detected transitions would greatly reduce the line-identification ambiguities affecting current blind searches. This concept is illustrated in Figs.~\ref{fig:3dtomography} and \ref{fig:tomographyspectra}. An illustrative 10-min ALMA2040 ``snapshot'' recovers the primary target, an $L_{\rm IR}=3\times10^{12}\,L_\odot$ galaxy at $z=3.1$, together with foreground and background galaxies at $z=1.1$ and $z=4.3$, respectively. Using the COSMOS2025 catalogue \citep{Shuntov2025}, we estimate that 64\% of galaxies at $z=2–4$ have at least one neighbour with SFR$>10$ $M_{\sun}$ yr$^{-1}$ within $7.5''$ out to $z<7$. Accounting for the snapshot sensitivity and frequency-dependent primary-beam attenuation, we find that 83\% (47\%) of such neighbours would be detected in $\geq1$ ($\geq2$) line at $\geq5\sigma$.

For the target itself, as well as neighboring sources, the spectrum would no longer contain only one deliberately selected transition (Fig.~\ref{fig:tomographyspectra}). Depending on redshift, ALMA2040 could simultaneously detect multiple CO transitions, [C\,I], H$_2$O, OH, OH$^+$, CH$^+$, HCN, HCO$^+$, and other species. At higher redshift, [C\,II] 158~$\mu$m, [N\,II] 205 and 122~$\mu$m, [O\,I] 145 and 63~$\mu$m, and [O\,III] 88 and 52~$\mu$m progressively enter the ALMA windows. Rather than deriving $M_{\rm mol}$ from a single CO transition using assumed excitation and $\alpha_{\rm CO}$, these observations would enable joint modeling of the molecular ISM. Multi-transition radiative-transfer approaches, including LVG/PDR models and frameworks such as \textsc{tuner}, can jointly model CO excitation, [C\,I], and dust continuum to constrain molecular gas mass, density, kinetic temperature, radiation field, and optical depth \citep{Harrington2021, Boogaard2026}. Thus, a single short ALMA2040 observation would move molecular-gas studies from single-tracer measurements to multi-line, multi-species ISM inventories for targets and serendipitous galaxies. Thousands of independent pointings would collectively form the distributed tier, complementing the contiguous surveys.


\subsection{Tracing the baryon cycle in action}

The combination of broad spectral coverage, high sensitivity, and wide-area surveys will allow ALMA2040 to move beyond inventories of molecular gas and directly trace its consumption and transport. H$_2$O emission, which correlates closely with far-IR luminosity, can provide a complementary spectroscopic tracer of obscured star formation \citep[e.g.,][]{Yang2019}. Meanwhile, OH, OH$^+$ or CH$^+$ probe molecular and diffuse gas flows, revealing inflowing and outflowing gas, or turbulent material through absorption, broad wings, and kinematically offset components \citep{Falgarone2017, Spilker2020, Riechers2021, Butler2023, Salak2024}. These diagnostics are currently accessible primarily in luminous or magnified systems. ALMA2040 could extend them toward broader galaxy populations. Combined measurements of CO, [C\,I], H$_2$O, OH, OH$^+$ and CH$^+$ would connect $M_{\rm mol}$, SFR, and gas flows within the same observations, constraining depletion times and mass-loading factors.

Wide, untargeted surveys will extend this experiment beyond individual galaxies, providing opportunities to catch the baryon cycle ``in the act.'' Spatially and kinematically offset cold-gas reservoirs, extended [C\,II] emission around dense protoclusters \citep[e.g.,][]{Solimano2024, Zhou2025, Harrington2025}, and [C\,I] potentially tracing accreting material \citep{Emonts2023} already suggest that gas transport can be identified through three-dimensional spatial and velocity information. Combined with wide-area optical and near-IR surveys, ALMA2040 will systematically search for such signatures across galaxies and environments. Angular resolutions of $\sim0.5$--$1.0\arcsec$ will be sufficient to identify spatial and kinematic offsets at high frequencies while retaining the surface-brightness sensitivity required for extended emission. These observations will identify candidate signatures of gas delivery to galaxies, molecular-gas consumption, feedback-driven ejection, and redistribution within groups and protoclusters. Joint spatial and kinematic analysis with stellar and ionized-gas tracers will help distinguish these processes and test whether displaced material is associated with companions, outflows, or accretion.

\subsection{Dust and discovery space}

The spectral-line inventory would be accompanied by an equally powerful continuum experiment. Broad multi-band coverage would densely sample the rest-frame far-IR/submillimeter SED of high-redshift galaxies (Fig.~\ref{fig:tomographyspectra}), replacing the one or two Rayleigh--Jeans measurements typical of current ALMA observations. This would improve constraints on $L_{\rm IR}$, dust temperature, emissivity, and dust mass while reducing the temperature--mass degeneracy. The same surveys could therefore measure $\rho_{\rm H_2}(z)$, the obscured cosmic SFR density, and the cosmic dust-mass density over matched volumes, connecting the evolution of molecular gas, star formation, dust, and metals.

Large-volume molecular surveys will also uncover systems that depart from established scaling relations. ASPECS has already identified gas-rich galaxies below the main sequence (Sect.~1.2), but current samples cannot establish their incidence or evolution, while wider surveys are beginning to uncover additional rare populations \citep{Catan2026}. ALMA2040 would measure the demographics and search for molecular gas-rich passive galaxies, low-mass systems, extreme gas fractions and dust-to-gas ratios, unusual excitation, strong gas flows, and sources without obvious optical/NIR counterparts. Finding such outliers is a key benefit of moving beyond samples selected where molecular-gas detections are already expected.


\begin{figure}[t]
\centering
\includegraphics[width=0.55\linewidth]{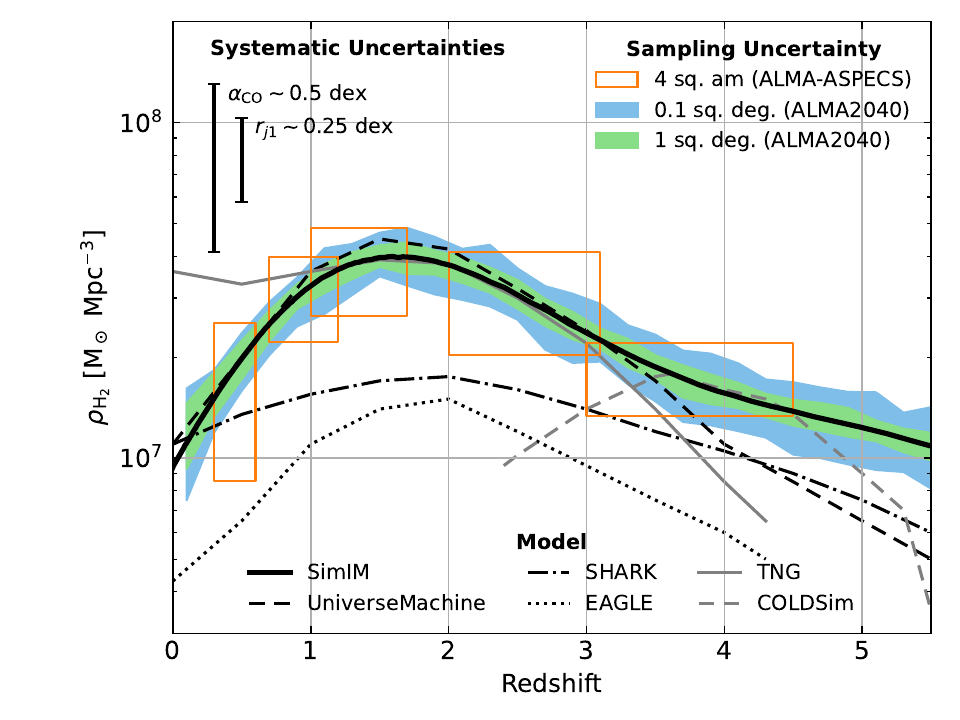}
\caption{
Cosmic molecular gas density, $\rho({\rm H}_2)$, as a function of redshift. The thick black curve (``SimIM'') shows a revised version of \citet{Keenan2020}. Blue and green regions show the sampling uncertainty from 100 mock light cones covering 0.1 and 1.0~deg$^2$, respectively, in $\Delta z=0.2$ bins; orange boxes show the corresponding uncertainty for ASPECS-sized ($\sim4$~arcmin$^2$) fields using the redshift bins of current CO deep-field measurements. Additional model predictions are compiled from \citet{Bollo2025}. 
The figure illustrates how wide-area surveys dramatically reduce sampling uncertainty, leaving molecular-gas conversion factors as an increasingly important systematic limitation. Although error bars illustrate representative systematic uncertainties for $\alpha_{\rm CO}$ and CO excitation \citep[e.g.,][]{Keenan2024}, ALMA2040 will constrain excitation and reduce these systematic uncertainties. 
}\label{fig:cosmicdensity}
\end{figure}


\section{Synergies}

\noindent {\bf Synergies with galaxy formation models.}~Current CO censuses reveal tensions in gas-rich galaxy abundances and
$\rho_{\mathrm{H}_2}$ evolution \citep[Fig.~\ref{fig:cosmicdensity}; e.g.,][]{Popping2019, Popping2022}. Larger samples and multi-tracer constraints will help distinguish model discrepancies from cosmic variance, uncertain gas conversions, and the sub-resolution ISM prescriptions used to predict line emission.
The wide and distributed ALMA2040 tiers could reach $\sim0.1$~dex sampling precision in $\rho_{\rm H_2}$, helping discriminate among model predictions while residual gas-conversion uncertainties are assessed separately.\\



\noindent {\bf Closing the baryon cycle.} By the 2040s, existing and forthcoming multiwavelength surveys will provide the stellar and environmental context for the ALMA2040 gas census. Rubin optical imaging combined with Euclid and Roman near-IR imaging will constrain photometric redshifts, stellar populations, and environments across fields suitable for the wide and distributed tiers. Deep JWST legacy fields and spectroscopy will provide complementary measurements of faint stellar populations, nebular attenuation, metallicity, and ionization conditions for the ultra-deep tier and selected sources. These measurements will help constrain metallicity-dependent gas-conversion factors and distinguish variations in excitation from changes in molecular gas content. 

ALMA2040 will add gas-selected redshifts, molecular reservoirs, and obscured star formation to populations characterized at optical and near-IR wavelengths, while also identifying sources with weak or absent counterparts. SKA and other radio facilities will constrain atomic gas out to $z\sim2$, and the ngVLA will provide complementary low-J CO measurements. Together, these datasets will connect gas fractions, depletion times, dust content, and gas flows to stellar mass, star formation, and environment.

\section{Technical requirements}

\subsection{Model of FIR/submm line emission at high redshift}
\label{sec:simulations}

We simulated 10-min ALMA2040 spectra across Bands~2--8 to illustrate
simultaneous ISM diagnostics and serendipitous detections
(see Figs.~\ref{fig:3dtomography} and ~\ref{fig:tomographyspectra}).


We adopted an empirical Arp~220 template using emission- and absorption-line measurements from \citet{Rangwala2011}, supplemented by modeled dense-gas transitions with \textsc{tuner} and representative far-IR fine-structure lines \citep{Decarli2025}. Lines were modeled as Gaussians with ${\rm FWHM}=400~{\rm km\,s^{-1}}$ and the continuum as a modified blackbody. The template, normalized to $L_{\rm IR}=10^{12}\,L_\odot$, was redshifted and rescaled to represent the main target (M), $(L_{\rm IR} [\,L_\odot],z, {\rm SFR} [M_{\sun} {\rm ~yr}^{-1}])$=$(3\times10^{12}, 3.1, 300)$; a background source (B), $(7\times10^{11}, 4.3, 70)$; and a foreground source (F), $(3\times10^{11}, 1.1, 30)$, sampled in $400~{\rm km\,s^{-1}}$ channels. The galaxies were placed at different positions in a mock ALMA field, attenuated by 12-m primary beam, convolved with a frequency-dependent synthesized beam (reference: $\sim2.1''$ at 100 GHz), and were assumed to be unresolved. The Arp220-based spectra illustrate accessible diagnostics for IR-luminous systems. Correlated Gaussian noise was added using prospective ALMA2040 sensitivities for a 10-min integration. The resulting cubes illustrate the simultaneous detection of continuum and multiple molecular, atomic, and fine-structure lines from both targeted and serendipitous galaxies (Figs.~\ref{fig:3dtomography} and \ref{fig:tomographyspectra}). Forecasts for the broader survey population require a range of excitation and dust-SED properties.

\subsection{Molecular deep fields}
\label{section:moleculardeepfields}
We envision a three-tier ``wedding-cake'' strategy for unbiased inventories of the molecular and dusty ISM. The wide tier retains ASPECS-like depth over a contiguous area far exceeding the $\sim50$--70~arcmin$^2$ scale at which cosmic variance and Poisson uncertainties become comparable, enabling measurements of large-scale structure with strongly reduced field-to-field variance.

We consider a 16,000-pointing hexagonal mosaic at 250~GHz. For a $\sim23\arcsec$ 12-m primary beam, a $15\arcsec$ spacing ($\sim1.3\times$ the nominal Nyquist spacing) provides an efficient compromise for simultaneous multi-band observations. The resulting $\sim0.24~{\rm deg^2}$ survey is nearly 190 times larger than the $4.6~{\rm arcmin^2}$ ASPECS LP and comparable to recent large ALMA surveys such as CHAMPS and exMORA \citep{Long2026, Faisst2026}. An ASPECS-like Band~6 sensitivity of $\sim0.5~{\rm mJy~beam^{-1}}$ per 18~km~s$^{-1}$ channel requires only $\sim20$~s per pointing with ALMA2040, or $\sim90$~hr on source for the full mosaic ($\sim180$~hr including current ALMA-like overheads). The survey would therefore reproduce ASPECS line sensitivity over nearly two orders of magnitude more area, overcoming the small-volume and cosmic-variance limitations of current deep fields. Band~7 is sparsely sampled ($\sim1.8\times$ Nyquist), but remains useful for bright line and continuum sources, whereas Bands~2--3 are oversampled and therefore reach greater effective depth. The same $\sim0.24~{\rm deg^2}$ field thus probes multiple tracers and redshift windows with progressively deeper coverage toward lower frequencies.
\begin{table}[t]
\centering
\caption{Predicted yields of the ALMA2040 tiers for simultaneous Band 2--8 coverage: $5\sigma$ line-flux limit at 230 GHz ($F_{5\sigma}$, Jy\,km\,s$^{-1}$ for 300 km/s linewidth), number of galaxies detected in $\geq1$ and $\geq2$ lines and at $z>4$, and number of CO and [C\,{\sc i}] line detections. The
[C\,{\sc i}] range reflects line ratios from the ASPECS carbon abundance (upper) and
line stacks \citep[lower;][]{Boogaard2020}.}
\label{tab:yields}
\small
\setlength{\tabcolsep}{4pt}
\begin{tabular}{lccccccc}
\hline
Tier & Area & $F_{5\sigma}$ & $N_{\rm gal}^{\geq1}$ & $N_{\rm gal}^{\geq2}$
 & $N_{\rm gal}^{z>4}$ & $N_{\rm CO}$ & $N_{\rm [CI]}$ \\
\hline
Wide       & 0.24\,deg$^2$   & 0.18  & $\sim$7000 & $\sim$6000 & $\sim$240 & $\sim$23\,000 & 4600--8600 \\
Ultra-deep & 4.5\,arcmin$^2$ & 0.014 & $\sim$160  & $\sim$145  & $\sim$15  & $\sim$700     & 150--220 \\
\hline
\end{tabular}
\end{table}
The ultra-deep tier would instead observe an ASPECS-sized field ($\sim4$--5~arcmin$^2$; $\sim80$ Band~6 pointings) for $\sim1$~hr per pointing, requiring $\sim80$~hr on source. It would probe substantially fainter ISM lines, constrain the faint end of the line-luminosity functions, and extend the molecular-gas census toward and beyond $z\gtrsim7$. Together with broad multi-line and continuum coverage, this provides the faint and high-redshift anchor of the wedding-cake strategy. The distributed tier would assemble approximately 5,000 independent extragalactic sightlines with at least 10 min on-source (see Fig. \ref{fig:tomographyspectra}), corresponding to $\sim$830 hrs of accumulated integration. Within the half-power primary beam, these pointings cover approximately 1.0 deg$^2$ at 100 GHz and 0.16 deg$^2$ at 250 GHz. Much of this integration could be accumulated through other science programs, although the yield depends on the number of unique suitable fields, their integration times, and their available spectral coverage.


We forecast yields from the ASPECS CO luminosity functions (LFs; Decarli et al. 2019, 2020): Schechter fits ($\alpha=-0.2$) at $z\approx0.5$--3.8, converted to CO(1--0) using observed excitation ratios \citep{Boogaard2020}, interpolated in $z$, and scaled outside this range following the cosmic SFR history \citep{Madau2014}. [C\,{\sc i}] assumes $X_{\rm [CI]}/X_{\rm H_2}=1.9\times10^{-5}$ and $T_{\rm ex}=30$\,K \citep{Boogaard2020}. A galaxy is detected in $\geq N$ lines when $L'_{\rm CO(1-0)}$ exceeds the $N$-th lowest excitation-corrected $5\sigma$ threshold among the CO and [C\,{\sc i}] lines within 67--500\,GHz. Applied to ASPECS, this reproduces the 1.2\,mm line candidates with fidelity $>0.8$ to $\sim$20\%; we thus scale yields by 0.8 and estimate uncertainties $\sim$30\%. Uncertainties are larger for the ultra-deep tier, which extrapolates the luminosity functions $\sim$1 dex below the ASPECS constraints.
We adopt $\alpha_{\rm CO}=3.6$, conversions to SFR as in \citet{Daddi2010}, and the [C\,{\sc ii}]--SFR calibration of \citet{DeLooze2014}. The wide tier would increase the ASPECS sample by over two orders of magnitude, with $\sim$85\% of galaxies detected in $\geq2$ lines, substantially reducing redshift ambiguities (Table~\ref{tab:yields}). At $z=2.5$, its CO(3--2) limit,
$L'\approx6\times10^{9}$\,K\,km\,s$^{-1}$\,pc$^2$ ($M_{\rm H_2}\approx2.6\times10^{10} \,M_\odot$, SFR\,$\approx35\,M_\odot$\,yr$^{-1}$), matches ASPECS, whereas the ultra-deep tier reaches $L'\approx5\times10^{8}$\,K\,km\,s$^{-1}$\,pc$^2$ ($M_{\rm H_2}\approx2\times10^{9}\,M_\odot$, SFR\,$\approx1.3\,M_\odot$\,yr$^{-1}$), a depth currently attainable only behind lensing clusters with $\mu\approx10$--15. For [C\,{\sc ii}], the ultra-deep tier reaches $L_{\rm [CII]}\approx0.9\,(1.6)\times 10^{7}\,L_\odot$ at $z\approx4.5\,(7)$, i.e., SFR\,$\approx0.7\,(1.4)\,M_\odot$\,yr$^{-1}$, $\sim10\times$ deeper than the wide tier and than targeted surveys such as ALPINE and REBELS \citep{Bethermin2020, Bouwens2022}.

\section{Summary}

Current surveys established the rise of $\rho_{\mathrm{H}_2}$ toward cosmic noon and its subsequent decline, but sensitivity, spectral coverage, and cosmic variance limit population-level measurements. ALMA2040 would combine a tenfold sensitivity gain, corresponding to 100 times faster integration at fixed depth, with broad simultaneous
spectral coverage. An ultra-deep $\sim4$ -$5\,\mathrm{arcmin}^{2}$ field, a wide $\sim0.24\,\mathrm{deg}^{2}$ mosaic, and thousands of independent sightlines would extend the census toward faint populations while sampling environments and large volumes. Multi-line observations would jointly constrain molecular reservoirs, excitation, dust, obscured star formation, and candidate gas flows. Together with multiwavelength surveys, these measurements would connect the gas census to stellar mass and environment, test galaxy formation models, and provide a population-resolved view of the baryon cycle.


\section*{Acknowledgements}
We thank the anonymous referees for helping improve this manuscript. MA is supported by FONDECYT grant number 1252054, and gratefully acknowledges support from ANID Basal Project FB210003,  ANID MILENIO NCN2024$\_$112 and ANID + Vinculaci\'on Internacional + FOVI250261. This work was developed as part of the broader ALMA2040 community effort. We acknowledge the Lorentz Center and NOVA for supporting the community workshops and coordination activities that helped initiate and shape the ALMA2040 White Paper process.\\

\bibliographystyle{apj_mod}
{\bibliography{bibliography}}

\printaffiliations
\end{document}